# The Carbon State in Dilute Germanium Carbides

I. A. Gulyas,[1] C. A. Stephenson,[2] Qian Meng,[3] S. R. Bank,[3]
M. A. Wistey[1]

[1]*Department of Physics, Texas State University, San Marcos, Texas 78666, USA*
[2]*Now at Sandia National Laboratory, Albuquerque, NM 87185, USA*
[3]*University of Texas at Austin, Austin, TX, 78705, USA*

Conduction and valence band states for the highly mismatched alloy (HMA) Ge:C are projected onto Ge crystal states, Ge vacancy states, and Ge/C atomic orbitals, revealing that substitutional carbon not only creates a direct bandgap, but the new conduction band is optically active. Overlap integrals of the new Ge:C conduction band with bands of pure Ge shows the new band has almost no Ge band character. C sites structurally mimic uncharged vacancies in the Ge lattice, similar to Hjalmarson's model for other HMAs. C perturbs the entire Ge band structure even at the deepest crystal core energy levels. Projection onto atomic sites shows relatively weak localization compared with other HMAs, but does show a strong anisotropy in probability distribution. L-valley conduction band states in Ge are ruled out as major contributors to the carbon state in Ge:C, both by weak inner products between these states and by a negligible effect on optical transition strength when adding C.

**I. INTRODUCTION**

Dilute germanium carbides (Ge:C) are a highly promising candidate for direct bandgap tunneling[1] and photonic[2,3] devices on silicon substrates,[4] with growth techniques that avoid undesirable carbon-carbon bonds during growth.[5,6,7,8] Highly mismatched alloys (HMAs) such as Ge:C, $GaAs_{1-x}N_x$, and $GaAs_{1-x}B_x$ exhibit properties beyond the ranges predicted by the virtual crystal approximation.[7,9,10] This opens new wavelengths for lasers,[11,12] detectors, and solar cells[13] while remaining lattice matched or at least compatible with Si, Ge, GaAs, or InP. Similarly, independent control of material properties could greatly improve steep switching in tunneling field effect transistors. Most semiconductors are constrained by tight coupling between material properties such as bandgap, lattice constant, and effective mass. But HMAs provide additional degrees of freedom even at small compositions of the mismatched atom, typically < 3 at.%. For example, the addition of only 2% nitrogen to InGaAs reduces the GaAs bandgap by more than 30% while simultaneously reducing strain on GaAs substrates.[14] At the other extreme, adding small amounts of B to GaAs appears to leave critical points of the band structure almost unchanged while reducing the lattice constant considerably.[15]

The mismatch in HMAs comes from the alloying of elements with the same valence but very different size, bond angle, and/or electronegativity. Such constraints typically reduce solubility limits to much less than 1% of the mismatched atom. This makes most HMAs difficult to synthesize except in very dilute quantities. Even kinetically limited techniques such as conventional molecular beam epitaxy may be unable to reach compositions approaching 1%.[16] The mismatch also makes HMAs a challenge to simulate numerically because the approximations that are frequently used for one atom species may be invalid for the other. For example, computational models of BN can ignore core electrons (none) and often omit spin-orbit coupling, but the "hard" or rapidly-varying potential near the N nucleus requires a large basis set of plane waves to accurately represent wavefunctions. On the other hand, "soft" InAs can be modeled with a smaller basis set, but inner *d* electrons and a relativistic Hamiltonian play a significant role. Strictly speaking, an accurate computational model of a HMA such as $InAs_{1-x}N_x$ therefore requires the combination of large supercells, many electrons per atom including inner electrons, a large basis set of plane waves, and spin-orbit coupling, resulting in an unfeasible computational demand.

To make such calculations tractable, GW many-body perturbation techniques have been used to approximately calculate energies of the isoelectronic carbon impurity or "defect" in Ge,[17] but large supercells were previously considered too computationally expensive for such techniques.[18] Unfortunately, the small bandgap of Ge leads to a metallic band structure when using supercells smaller than 64 atoms (1.6 at.% C). Even larger supercells are necessary if simulating the smaller carbon concentrations corresponding for bandgaps used for photonic integrated circuits or datacom, necessitating 128 atom supercells (0.78 at.% C).

Previous reports explored band structures of Ge:C and other HMAs. In dilute nitrides such as GaInAs$_{1-y}$N$_y$, band anticrossing (BAC) provides a satisfying, if somewhat simplified, prediction of band structure, which is readily modeled using k•p perturbation methods with a single additional band. Nitrogen creates a state within the conduction band that behaves atypically with pressure, as if it were pinned to the vacuum level rather than to the conduction band edge. This behavior is usually described using the deep trap model of Hjalmarson & Dow, in which the defect state is dominated by dangling bonds in the host.[19,20] In this context, "deep" is not "deep within the bandgap," but it refers to states whose origins are deep core atomic states and/or whose behavior is tied to the vacuum level, rather than the conduction or valence bands.

In light of recent reports that the lowest conduction band in Ge:C is optically inactive,[21] i.e. a pseudo-direct bandgap,[22] and after establishing suitable convergence conditions for accurate modeling of Ge:C,[23] we investigated the nature of the carbon states in Ge:C. In particular, we examined whether the electronic and optical properties of Ge:C, and their pressure dependence (or lack thereof)[20] can be explained using Hjalmarson's model. We applied computational techniques with higher accuracy than previously available. Although BAC correctly predicts a splitting of the conduction band at $k = 0$, it fails to describe the band structure at higher values of $k$, in contrast with many other HMAs such as GaAs$_{1-x}$N$_x$.

In this report, we use dopant notation (Ge:C) since the fraction of C, 0.78%, is quite small for an alloy. However, its effects are far more pronounced than typical dopants.

## II. COMPUTATIONAL METHODS

The Vienna *ab initio* Simulation Package (VASP)[24-27] was used to perform density functional theory (DFT) based simulations of a 128-site diamond lattice supercell of Ge, Ge:C, or Ge with a vacancy (v$_{Ge}$). For v$_{Ge}$, one Ge atom was removed; for Ge:C, it was replaced with C. The projector-augmented wave (PAW) core electron method was used with the generalized gradient approximation (GGA) and Perdew-Burke-Ernzerhof (PBE) functional.[28-31] PBE functionals are well known to underestimate bandgaps, but all-electron models are prohibitive due to the sheer numbers of atoms (128) and electrons (32 per Ge atom). Given $N$ total electrons, the complexity varies from $O(N^2 \ln N)$ for FFT-limited techniques to $O(N^3)$ if exactly diagonalizing the Hamiltonian. As a compromise, the Heyd-Scuseria-Ernzerhof (HSE06) hybrid functional was used instead of PBE after the first ionic relaxation.[32] Results from HSE06 and a similar functional, HSEsol, have been reported to be very similar.[33] Due to computational limits and the large supercells used here, inclusion of PBE0 hybrid functionals, spin-orbit-coupling, and $d$ electrons were impractical. Although these affect the valence band,[34] they were previously found to induce relatively minor changes in Ge conduction band structure, both in calculation[5] and theory.[35] Further simulation details can be found in Ref. 23.

The computational lattice constant was calculated by rigidly varying the supercell lattice vectors, relaxing the ion positions within each new volume, and fitting the resulting set of system energies to the Birch-Murnaghan equation of state. Computational lattice constants were used as-is, without attempting to force a fit to experimental parameters such as direct or indirect bandgaps. This had minimal effect on the results presented here, as verified by examining strained supercells (not shown).

In addition to hybrid functionals, we increased the number of k-points and plane waves beyond the default values for typical simulations, in order to more accurately capture short-range electronic structure. Specifically, we used a Γ-centered 2×2×2 or 3×3×3 mesh of k-points, which would be comparable to 8×8×8 or 12×12×12, respectively, in the primitive cell. Unless otherwise noted, an energy cutoff of 550 eV was used for plane wave basis sets, as this was found to be sufficient for good convergence.[23] To accurately capture higher conduction band states, free carrier absorption, and effective masses, unless otherwise indicated, 784 bands were included in the calculation, many of which were degenerate, particularly at higher levels. For comparison, the 256$^{th}$ band was the last filled valence band, and the rest were conduction bands; additional bands better captured the character of upper conduction band states, and therefore higher optical transitions such as free carrier absorption (FCA).

VASP normalizes its wavefunctions using an overlap operator rather than directly setting the norm $\langle \psi_n^* | \psi_n \rangle = 1$. The difference is typically relatively small. For example, we found that a 2-atom GaAs cell with 56 bands showed a maximum norm $\langle \psi_n^* | \psi_n \rangle$ of 1.45, a minimum of 0.894, and a standard deviation of 0.099. Similarly, for 128-atom Ge:C supercells with 336 bands presented here, the range of norms was 1.04-1.37 with a standard deviation of 0.083. However, in order to accurately compare even small differences in

matrix elements between two states, the wavefunctions of the two states were normalized using the components of the wavefunction as follows. Given a plane wave basis set at a given $k$ and truncated at $|hk| \leq E_{max}$:

$$\psi_k(\vec{r}) = \sum_{n=0}^{\frac{E_{max}}{\hbar}} a_{n,k} e^{-in\vec{k}\cdot\vec{r}} \quad (1)$$

The inner product or overlap integral between states $\psi_i$ and $\psi_f$ is

$$\langle \psi_{i,k}^* | \psi_{f,k} \rangle = \sum_{n=0}^{\frac{E_{max}}{\hbar}} a_{n,k}^* b_{n,k} \quad (2)$$

where $a$ and $b$ are the coefficients of the plane waves forming the initial and final states, respectively. These plane wave coefficients were extracted from the WAVECAR file using WaveTrans.[36] For the calculations reported here, the plane wave coefficients returned by VASP at each value of $\vec{k}$ were independently rescaled for a norm of 1, i.e.

$$\langle \psi_{i,k}^* | \psi_{f,k} \rangle = 1 \text{ if } i = f \quad (3)$$

Eqn. (2) was used to project one state onto another, such as when quantifying the similarity between the C states in Ge:C and the original unmodified conduction band states in Ge. Similarly, the relative transition strength or matrix element $M_{i,f}$ of optical transitions between two states $\psi_{i,k}$ and $\psi_{f,k}$ is determined from the momentum operator and Fermi's Golden Rule:

$$M_{i,f} = \langle \psi_{i,k}^*(\vec{r}) | H'(\vec{r}) | \psi_{f,k}(\vec{r}) \rangle$$
$$\propto \langle \psi_{i,k}^*(\vec{r}) | -i\hbar \hat{e} \cdot \nabla | \psi_{f,k}(\vec{r}) \rangle \quad (4)$$

where $\hat{e}$ is the unit vector of the optical electric field, which we assume from here on to be polarized such that the dot product is 1. Matrix elements from VASP were scaled using the same renormalization factors as in Eqn. (3).

In this work, the HSE screening parameter was slightly modified from 0.20 (HSE06) to 0.18 for comparison with previous work,[5] but computational lattice constants were not adjusted. Forcing a smaller lattice constant to fit both direct and indirect bandgaps would not directly affect the main points of this work, although it could influence whether a given alloy and strain are direct or indirect if the difference in energies is small. Also, in the absence of both spin-orbit coupling and strain, the three valence bands are all degenerate at Γ.

When the crystal primitive cell is repeated to form a supercell, the X and L points in the first Brillouin zone get folded over onto the Γ point. For the 4×4×4 cells used here, this folding of the Brillouin zone occurs twice. Thus, the energy eigenvalues at "$k$=0" actually contain the union of the band energies for $k$ = <000>, <100>, <111>, ½<100>, ½<111>, ¼<100>, and ¼<111>, making interpretation nontrivial. To identify the equivalent $k$ in the primitive cell Brillouin zone, either the bands were unfolded using either BandUP[37-40] or vasp_unfold,[41] or the character of the band was determined manually from the projection onto the $s$ and $p$ orbitals, which is also the method used by vasp_unfold. Because each state is divided among 128 atoms, to reduce rounding errors when projecting onto atomic orbitals, VASP was modified to produce 6 digits of precision in its PROCAR output files. Also, due to the periodic boundary conditions, it does not matter which Ge atom is replaced by C, since the overall periodicity and resulting band structures would be identical.

## III. RESULTS AND DISCUSSION

Adding dilute C to Ge has previously been shown to decrease the bandgap at Γ, as in dilute nitrides. The bottom conduction band (CB) is split into E$^+$ and E$^-$ bands at Γ, along with an *increase* in the conduction effective mass, and these are predicted by the BAC model. But the effects of C on the band structure are not well captured by BAC away from $k$=0, with a plateau in the CB toward L and a rise toward X. What are the band and bond origins of the carbon state in Ge, and why does it show such asymmetry in the band structure?

Fig. 1 shows the projection onto atomic orbitals of the Γ states at the valence band (VB) maximum, which is triply degenerate, as well as the lowest CB minima. 128-atom supercells of Ge, Ge$_{127}$C$_1$, and Ge$_{127}$v$_1$ were used with a 2×2×2 mesh of k-points. The plane wave basis set included waves up to 520, 550, and 600 eV (ENCUT parameter) for Ge, Ge:C, and v$_{Ge}$, respectively, and each was simulated at its own relaxed lattice constant. Contrary to recent reports,[21] both of the split CBs in Ge:C are predominantly $s$-like in character over most atoms, like the unperturbed Ge CB, and the VBs likewise retain their predominantly $p$-like character. This means both the E+ and E- CBs in Ge:C will be optically active for band-to-band transitions.

The similar *s*-like nature of the split E+/E- CBs does match the premise and predictions of the band anticrossing model at Γ, although we shall show later that the first order BAC model rapidly fails to explain band structures away from k=0. Adding a vacancy in Ge ($v_{Ge}$) introduces a new, empty *p*-like state 0.65 eV above the VB edge and a mostly *s*-like state above the CB edge, tentatively identified as $E_{a1}$ and $E_{d1}$, respectively. The Ge:C E+/E- bands appear to have very different character from the vacancy state, though both C and vacancies are treated in Hjalmarson's model as "deep" states.[19] However, we shall see later that other similarities do exist. We note in passing that the $E_{vc}$ state is triply degenerate.

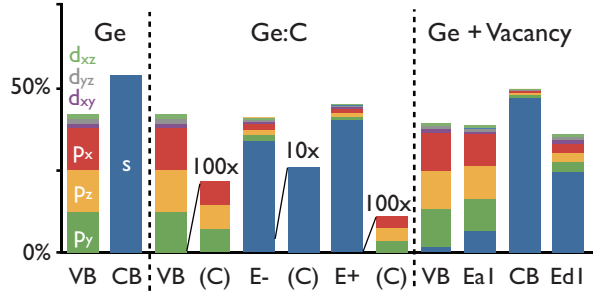

Fig. 1. Atomic orbital character of states near bandgap at Γ. The projections onto orbitals are summed over all atoms, except projections onto the C ion alone are also shown for Ge:C. Degenerate bands are averaged. Totals are less than 100% because only outer orbitals and contributions within the Wigner–Seitz radius of each atom are included.

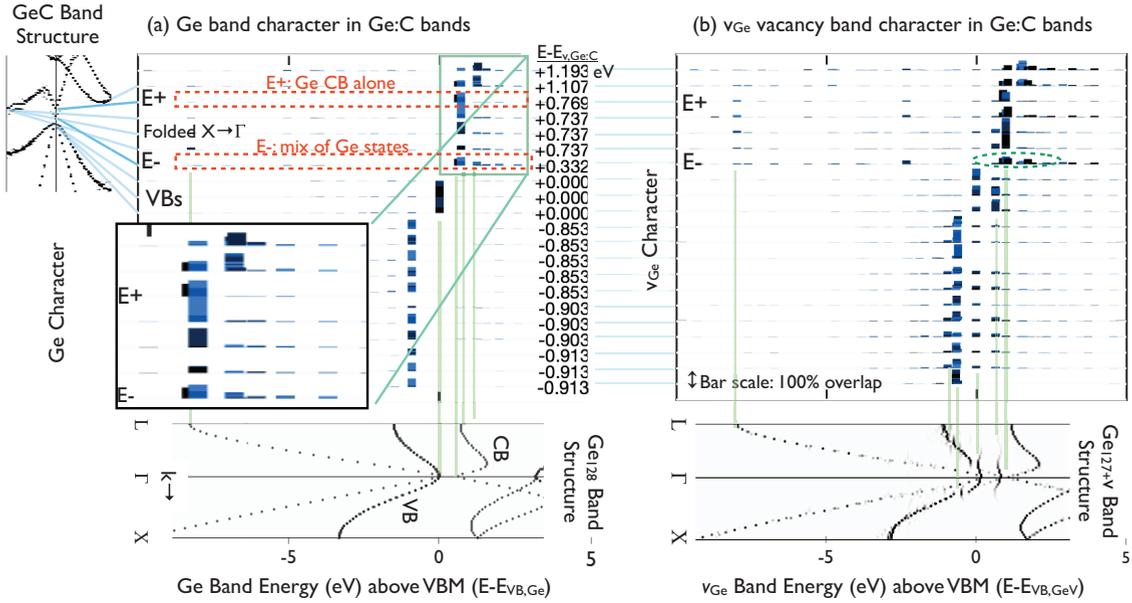

Fig. 2. Projection of several Γ and Γ-folded Ge:C states onto states from either (a) pure $Ge_{128}$ or (b) Ge with a single vacancy. Length of each bar represents the inner product between the given Ge:C state and the Ge or $v_{Ge}$ state at the given energy. For clarity, bars from closely-spaced bands are stacked, and GeC bands are offset vertically, labeled as energy (eV) above Ge:C VBM. Straight blue and green lines are guides to the corresponding points in respective band structures.

Although the projection onto atomic orbitals provides a qualitative measure of localization on the C atom, it may lose important quantitative information about the character of the state within the band, e.g.

optical or transport properties. To address this, we calculated the inner product of the E- state (and the E+ state) with unperturbed Ge wavefunctions that were simulated using the same basis set and conditions as Ge:C. This allows the identification of which bands in the Ge crystal the alloy borrowed from or mixed with to create the carbon state(s).

In order to identify how the C states gain their character, including pressure dependence, states near the the Ge:C band edge were compared with those from either pure Ge or Ge with a single vacancy ($v_{Ge}$). As shown in Fig. 2, the Ge:C valence bands and *upper* conduction bands are seen to share character with the corresponding bands in Ge. But the lower conduction band, E-, is not well modeled by any single Ge band. Rather, E- is a mix of many Ge bands at Γ. By comparison, E- has a significant projection onto the defect state, as circled in Fig. 2(b). Repeated similar features in Fig. 2 are due to the degeneracy of the three valence bands as well as the upper conduction bands in this system. The presence of multiple bars per band indicates mixing from different states across a range of energies. Higher CBs also have some low-VB $v_{Ge}$ character (dashed circle). The $v_{Ge}$ band structure in Fig. 2 was calculated using only 504 bands. However, band energies and transitions at k=0 were calculated using 784 bands as mentioned above. Additional similarities between the C "defect" and a Ge vacancy are shown in the filled-state charge density, as shown in Fig. 3. In $v_{Ge}$ and Ge:C, the defects clearly affect longer range charge densities, at least to the 3$^{rd}$-nearest neighboring atoms. Surprisingly, although $v_{Ge}$ and Ge:C show nearly identical charge distributions away from the defect, Ge:C shows significant charge strongly localized on the C atom. Since this is a filled-state plot, it shows that the charge on C must arise from valence states (i.e. filled states). Additionally, since the charge is strongly localized, it is also strongly bonded, arising from states at low energies deep within the valence bands.

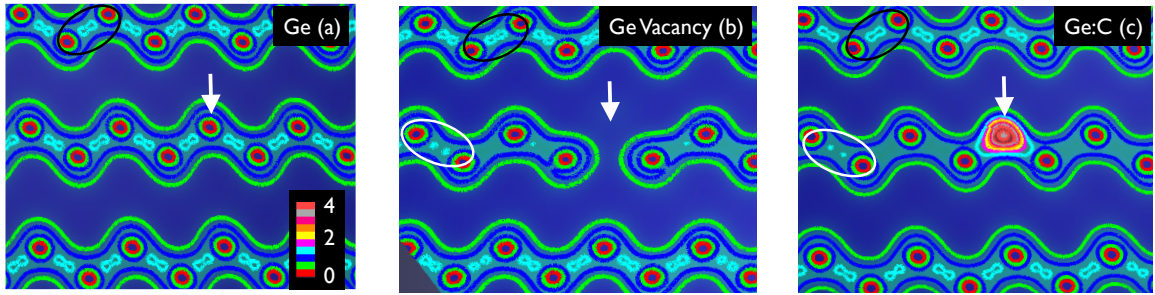

Fig. 3. Filled state charge density contour plots along a {110} plane. (a) Unperturbed Ge, (b) Ge with a single vacancy, (c) Ge:C. Equivalent positions are marked with white arrows. Although a C atom attracts charge and the vacancy repels it, both types of defect reduce the charge in bonds as far away as 3rd-nearest neighbors along <110> (white ovals) compared with other directions (black ovals) or any bond in unperturbed Ge. Color scales are identical.

The E- band at Γ was reported by Kirwan et al. to be independent of pressure, which was attributed to a mixing of L states into the band at Γ. This would lead to a pseudo-direct bandgap[22] (not to be confused with a nearly-direct bandgap) in which the lowest conduction band states had the wrong symmetry for strong optical transitions to or from the valence band. As mentioned above, optical transitions impose a momentum operator in the inner product between the initial and final states, so a symmetric final state becomes antisymmetric in the integral, and vice versa. Therefore, the strongest transitions are those from symmetric to antisymmetric states, or vice versa, such as between *s* and *p* states. To examine whether this independence was due to L-valley states or deep *s*-like states instead, we extracted optical transition matrix elements for transitions from the valence bands to the E- band. The projection onto atomic orbitals in Fig. 1 verified that the valence bands were still 90% *p* orbitals, which are antisymmetric with respect to the atom core. Electron states at L are likewise antisymmetric due to the Bloch waves at the edge of the Brillouin zone, so addition of L states at Γ would tend to reduce the strength of optical transitions. Instead, we found that the strength of the optical transition from VB maximum to CB minimum (E-) in Ge:C <u>was within 6% of that in Ge. The</u>

VB-E$^+$ transition was within a factor of 2 of Ge, but all other nearby transitions were two orders of magnitude smaller. This supports the symmetric, s-like symmetry of the CB. Optical transitions will be reported in more detail elsewhere. The strong optical transitions to the E$^-$ band mean that the reported pressure independence comes from a source other than indirect L-valley states, again consistent with Hjalmarson's model.

Table I. Structural shifts due to C atom or Ge vacancy.

|  | Nearest atom shift toward defect (pm) | Distance between neighbors closest to defect (pm) | Angle through neighbor to defect (degrees) | Bond angle of 1$^{st}$- to 2$^{nd}$-nearest neighbors (degrees) |
|---|---|---|---|---|
| Ge (pure) | 0 | 404 | 109.5 (all bonds) | 109.5 (all bonds) |
| Ge vacancy | 29 | 355 | 115.1 | 103.3 |
| Ge:C | 35 | 345 | 115.6 | 102.8 |

The similarity between a Ge vacancy and a carbon atom in Ge becomes even more clear from Table I. Both types of defect shift the four Ge nearest neighbor atoms a comparable distance inward toward the defect, 12-14% of the original 247 pm bond length. The angle from second-nearest neighbors through the nearest neighbor to the defect is distorted from the original dihedral angle of 109.5° to roughly 103°, and the second-nearest bonds are conversely increased from 109.5° to above 115°.

A significant contrast between Ge:C and $v_{Ge}$ occurs in the top three VBs. A significant fraction of character of the top three Ge:C valence bands is shared with the $v_{Ge}$ acceptor-like state deep within the bandgap. But this appears to be the $v_{Ge}$ acceptor-like $E_{a1}$ state picking up VB character rather than the other way around; the top Ge:C VBs consist almost entirely of unperturbed Ge VBs, as shown by the 100% overlap near E=0. Because the $v_{Ge}$ acceptor-like state shares so much character with the Ge VB, this suggests BAC in the *valence* band of $v_{Ge}$, which we do not observe for Ge:C. However, the influence of C does extend to the bottom of the VB. Projection of the very deepest valence band state (~11 eV below the bandgap) onto individual atoms shows that it is 42% on the carbon atom, despite C being only 0.78% of the alloy. This C contribution is 100.00% *s*-like in character.

An exploratory look at a less-symmetric system initially appeared to rule out supercell periodicity as the origin for the charge anisotropy in Fig. 3. We modeled a cubic Ge$_{63}$C$_1$ system in which neighboring supercells were staggered with basis vectors (0.5, 0.75, 075), (0.625, 0.125, 0.5), (0.5, 0.5, 0), rather than meeting at face centers. As shown in Fig. 4, this had the effect of changing the distance and direction between C atoms along different directions. Although no anisotropy was apparent in charge density, there was also no visible difference along the zigzag bonds in the {110} planes. On the other hand, 64 atom Ge:C supercells have previously been found to produce a metallic band structure, which may explain the charge uniformity along the chains of bonds. In other words, the virtually identical Ge-Ge bond charge throughout the solid might not be due to a lack of interaction between neighboring C atoms, but rather due to electrons being shared *too well*, leading to a uniform charge distribution. Or alternately, the isotropic bonds might somehow be due to the lower-accuracy (higher speed) simulation conditions: VASP defaults for energy cutoff (ENCUT), energy convergence criterion (dE), precision (PREC = Normal), and 336 bands.

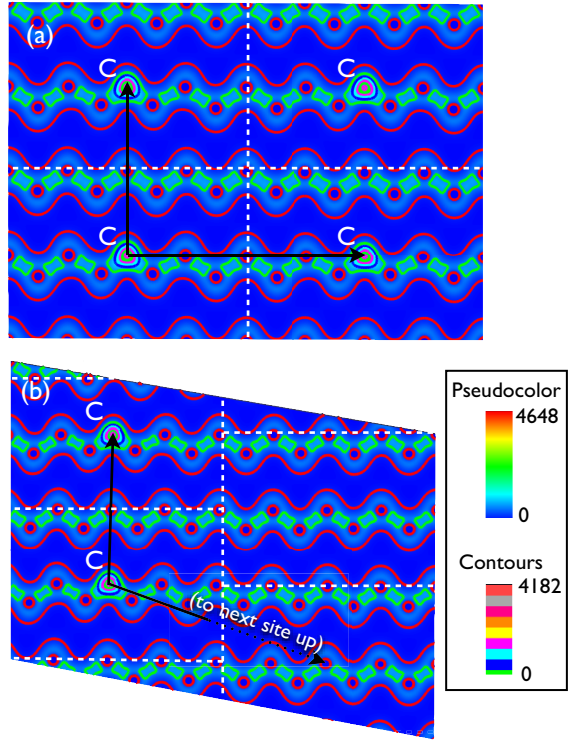

Fig. 4. Charge density slice along (110) including multiple adjacent, cubic, 64-atom supercells with different basis vectors (black arrows). (a) x/y/z basis aligned with 8-atom simple cubic cell, sliced through the C atom. (b) Neighboring supercells shifted by 1-2 sites perpendicular to plane. C atom only appears in alternating atomic planes of the supercell.

This leaves the question why BAC fails to represent the band structure away from $k$=0. For this, the E-band was projected onto the full set of atoms (Fig. 5(ab)), and the results summed by position projected into the x-y plane (Fig. 5(cd)). For conventional supercells with minimum length basis vectors, a weakly aligned charge density along <111> appears. This is also consistent with the 4$^{th}$-nearest neighbor effects seen in Fig. 3. Using staggered supercells, where the center of one meets a vertex or edge of the next, significant anisotropy appears along <111> directions.

Even though the carbon "defect" is localized, its nonlocal charge (at least the fraction projected onto atomic orbitals) extends 4 atoms away along <111> directions. The anisotropy shown in Fig. 5 is not unique to the Ge-C system, but it is somewhat more delocalized than other HMAs such as GaAs:N or GaP:N.[42,43,44] We attribute the difference to the reduced difference in electronegativity: $\chi_B$ - $\chi_{Ge}$ = 0.7, compared with $\chi_N$ – $\chi_{As}$ = 1.0.

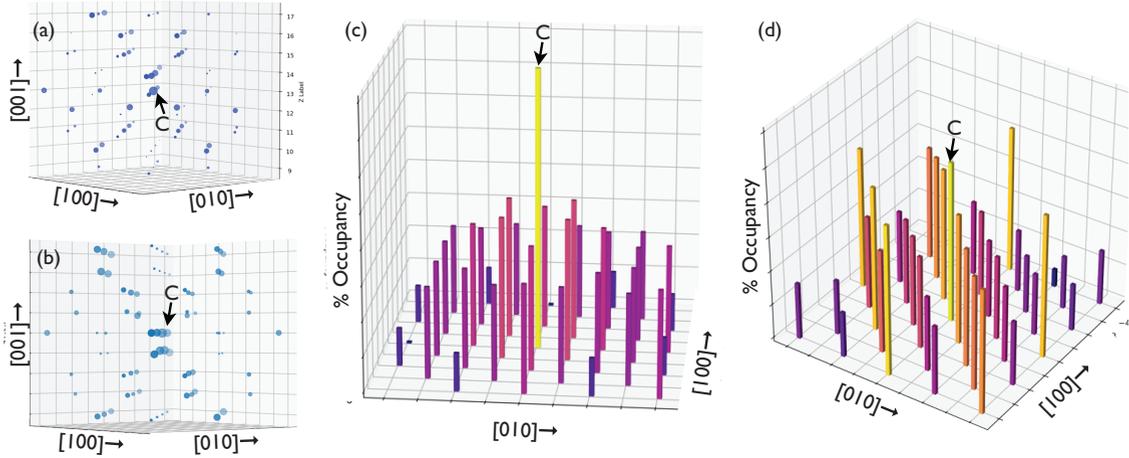

Fig. 5. Projection of lowest Ge:C CB (E-) states onto ions near C atom for 128 atom supercells repeated with different basis sets. (a),(b): 3D scatter plot. Spot diameter represents amount of wavefunction on atom at that location. Cube volume chosen to match a supercell, with 128 atoms. (c),(d): Sum of any vertically-aligned atoms in (a) and (b), respectively, projected down onto the x-y plane as bar height. Bar height represents the fraction of state localized on any atoms at that x-y location within the cube. (a) and (c) are from conventional supercell stacking (face to face); (b) and (d) are from staggered supercells similar to Fig. 4(b).

We interpret this <111> symmetry to be responsible for the divergence of band structures from the band anticrossing model when looking in different crystal directions. The difference in distribution of the E- band over different atoms changes distinctly with different supercell basis vectors, even though the C atoms are always at least 7 bonds away from each other. In light of this observation, it may be worth asking whether the anisotropic wavefunctions reported for similar HMAs are likewise due to choice of supercell basis vectors. However, it is also worth noting that Fig. 5 does not plot actual wavefunction probability density, but the projection onto atomic orbitals. It may be possible that the orbitals along <111> simply line up with the E- wavefunction better than others. Future work will study whether how such variations of a longer-ranged ordered alloy affect thermodynamic favorability, supercell shape, and band structure.

## IV. ADDITIONAL DISCUSSION

Optical transitions from *p*-like to *d*-like states are also possible, so it is worth examining whether the strong optical transitions to the E- band might instead be caused by a mixture of *d*-like states. Although *d* core electrons were not included in the PAW atomic *potentials* used here, the projection of final *wavefunctions* onto *d*-orbitals would still indicate whether a tendency toward *d*-orbitals was significant. However, projection of lowest conduction band onto Ge and Ge:C atomic orbitals shows very weak contribution from *d*-orbitals, <2%, as shown in Table II. The small fractions of *p* and *d* character in the E- band do not explain the strong optical transitions remaining after adding C. We therefore believe Ge:C to be a true direct bandgap with mostly *s*-like symmetry and strong optical transitions across the bandgap to and from the two lowest conduction bands. This is consistent with $GaAs_{1-x}N_x$; even though N constitutes only 1-1.5% of all atoms in the crystal, it increases the optical matrix element by a factor of 2.[43] Even if the fraction of C in the alloy was small enough that $E_{CB,L} < E_{CB,\Gamma}$, it would still be nearly-direct rather than pseudo-direct.

Table II. Projection of lowest Γ conduction band onto Ge and Ge:C orbitals, summed over all atoms. Conditions are same as Fig. 1.

| CB | s | $p_y$ | $p_z$ | $p_x$ | $d_{xy}$ | $d_{yz}$ | $d_{z^2}$ | $d_{xz}$ | $d_{x^2-y^2}$ | Total |
|---|---|---|---|---|---|---|---|---|---|---|
| Ge | 0.1369 | 0.0000 | 0.0000 | 0.0000 | 0.0000 | 0.0000 | 0.0000 | 0.0000 | 0.0000 | 0.1369 |
| Ge:C | 0.5429 | 0.0273 | 0.0273 | 0.0273 | 0.0097 | 0.0097 | 0.0009 | 0.0097 | 0.0009 | 0.6560 |

The use of pseudopotentials near atomic cores raises the question of accuracy of projections of one state wavefunction onto another, and similarly the optical matrix elements for transitions between states. This is because pseudopotentials replace the actual potential near the atomic core with a constant or slowly-varying potential in order to reduce computational demands. Because potentials are deepest near the atomic cores, the fraction of wavefunction near the cores may be non-negligible and oscillate strongly with position. This would reduce the accuracy of overlap integrals for the optical matrix elements. However, PAW potentials do reconstruct the exact valence wavefunction with all nodes near the cores.[28,29] This significantly increases the accuracy not only of band structures, but also of the optical matrix elements, i.e. the overlap integral between the wavefunction of the initial state and the derivative of the wavefunction of the final state.

Finally, we note that adding a vacancy strongly affects the valence bands, and adds a VB-like state within the bandgap. This study was unable to include SOC or *d*-electrons, which do affect the valence band. Therefore, comparisons of Ge:C with $v_{Ge}$ should be considered qualitative rather than quantitative.

## VI. SUMMARY

In conclusion, 128 atom supercells of Ge:C with 1 C atom were modeled using hybrid functionals in order to study the origins of the new bands of states introduced by carbon. L-valley conduction band states in Ge are ruled out as major components of the Γ carbon state in Ge:C by both a lack of change in the optical matrix elements across the bandgap at Γ and by a negligible projection of Ge:C states onto L-valley states. Rather, the similar pressure independence comes from the fact that both sets of states come from orbitals farther from the bandgap. This is shown to be similar to vacancies in the Ge host matrix, in accordance with Hjalmarson's deep state model. However, Ge:C is not well described by the band anticrossing model away from *k*=0. The differences from the first-order BAC model may be due to the bond distortion imposed by the introduction of C into the lattice, or by the strong and anisotropic charge in the E- band observed along <111> for some atom configurations, even with C atoms 8 or more bonds away from each other. Although this could not be distinguished from super-periodicity imposed by the use of finite supercells in the present study, it is noteworthy that the physical range of filled-state charge displacement and the E- CB state wavefunction both reach at least as far as 4$^{th}$-nearest neighboring atoms.

## ACKNOWLEDGMENTS

This work used the Extreme Science and Engineering Discovery Environment (XSEDE) through allocation DMR140133, supported by National Science Foundation grant number ACI-1548562. It was supported in part by the National Science Foundation under Grants DMR-1508646 and CBET-1438608, a Notre Dame Energy Center postdoctoral fellowship, the Notre Dame Center for Research Computing, and the Texas State University LEAP center. VisIt visualization software is supported by the Department of Energy with funding from the Advanced Simulation and Computing Program and the Scientific Discovery through Advanced Computing Program. The authors also thank Eoin P. O'Reilly for early access to Ref. 20.

## REFERENCES

1. I. A. Gulyas and M. A. Wistey, in preparation.

2. D. Gall, J. D'Arcy-Gall and J. Greene, Phys. Rev. B **62**, R7723 (2000).

3. R. Soref, Silicon Photonics VIII **8629**, 862902 (2013).

4. M. Wistey, Y.-Y. Fang, J. Tolle, A. G. Chizmeshya and J. Kouvetakis, Appl. Phys. Lett. **90** (8), 082108 (2007).

5. C. Stephenson, W. O'Brien, M. Penninger, W. Schneider, M. Gillett-Kunnath, J. Zajicek, K. Yu, R. Kudrawiec, R. Stillwell and M. Wistey, J. Appl. Phys. **120** (5), 053102 (2016).

6. J. Kouvetakis, A. Haaland, D. J. Shorokhov, H. V. Volden, G. V. Girichev, V. I. Sokolov and P. Matsunaga, J. Am. Chem. Soc. **120** (27), 6738-6744 (1998).

7. J. D'Arcy-Gall, D. Gall, I. Petrov, P. Desjardins and J. Greene, J. Appl. Phys. **90** (8), 3910-3918 (2001).

8. J. D'Arcy-Gall, D. Gall, P. Desjardins, I. Petrov and J. Greene, Phys. Rev. B **62** (16), 11203 (2000).

9. S. Zhang and S.-H. Wei, Phys. Rev. Lett. **86** (9), 1789 (2001).

10. R. H. El-Jaroudi, K. M. McNicholas, B. A. Bouslog, I. E. Olivares, R. C. White, J. A. McArthur and S. R. Bank, presented at the Conference on Lasers and Electro-Optics (CLEO), San Jose, California, 2019 (unpublished).

11. S. Bank, H. Bae, H. Yuen, M. Wistey, L. L. Goddard and J. Harris, Electron. Lett. **42** (3), 156-157 (2006).

12. M. Wistey, S. Bank, H. Bae, H. Yuen, E. Pickett, L. L. Goddard and J. Harris, Electron. Lett. **42** (5), 282-283 (2006).

13. D. B. Jackrel, S. R. Bank, H. B. Yuen, M. A. Wistey, J. S. Harris Jr, A. J. Ptak, S. W. Johnston, D. J. Friedman and S. R. Kurtz, J. Appl. Phys. **101** (11), 114916 (2007).

14. E. D. Jones, N. A. Modine, A. A. Allerman, I. J. Fritz, S. R. Kurtz, A. F. Wright, S. T. Tozer and X. Wei, presented at the Light-Emitting Diodes: Research, Manufacturing, and Applications III, 1999 (unpublished).

15. I. Gulyas, R. Kudrawiec and M. A. Wistey, J. Appl. Phys. **126** (9), 095703 (2019).

16. S. Park, J. D'Arcy-Gall, D. Gall, Y.-W. Kim, P. Desjardins and J. Greene, J. Appl. Phys. **91** (6), 3644-3652 (2002).

17. S. Ismail-Beigi and S. G. Louie, Phys. Rev. Lett. **95** (15), 156401 (2005).

18. H. Koyama and K. Sueoka, J. Crystal Growth **463**, 110-115 (2017).

19. H. P. Hjalmarson, P. Vogl, D. J. Wolford and J. D. Dow, Phys. Rev. Lett. **44** (12), 810 (1980).

20. P. Vogl, H. P. Hjalmarson and J. D. Dow, J. Phys. Chem. Solids **44** (5), 365-378 (1983).

21. A. C. Kirwan, S. Schulz and E. P. O'Reilly, Semicond. Sci. Technol. **34,** 075007 (2019).

22. J. Shay and J. Wernick, Ternary Chalcopyrite Semiconductors (Pergamon Press, Oxford, 1975).
23. I. A. Gulyas, C. A. Stephenson and M. A. Wistey, submitted to Comp. Mat. Sci..
24. G. Kresse and J. Hafner, Phys. Rev. B **47** (1), 558 (1993).
25. G. Kresse and J. Hafner, Phys. Rev. B **49** (20), 14251 (1994).
26. G. Kresse and J. Furthmüller, Comp. Mat. Sci. **6** (1), 15-50 (1996).
27. G. Kresse and J. Furthmüller, Phys. Rev. B **54** (16), 11169 (1996).
28. P. E. Blöchl, Phys. Rev. B **50** (24), 17953 (1994).
29. G. Kresse and D. Joubert, Phys. Rev. B **59** (3), 1758 (1999).
30. J. P. Perdew, K. Burke and M. Ernzerhof, Phys. Rev. Lett. **77** (18), 3865 (1996).
31. J. P. Perdew, K. Burke and M. Ernzerhof, Phys. Rev. Lett. **78** (7), 1396-1396 (1997).
32. A. V. Krukau, O. A. Vydrov, A. F. Izmaylov and G. E. Scuseria, J. Chem. Phys. **125** (22), 224106 (2006).
33. J. Hong, A. Stroppa, J. Íniguez, S. Picozzi and D. Vanderbilt, Phys. Rev. B **85** (5), 054417 (2012).
34. S. H. Wei and A. Zunger, J. Vac. Sci. Technol. B**5** (4), 1239-1245 (1987).
35. L.-D. Yuan, H.-X. Deng, S.-S. Li, S.-H. Wei and J.-W. Luo, Phys. Rev. B **98** (24), 245203 (2018).
36. R. M. Feenstra, N. Srivastava, Q. Gao, M. Widom, B. Diaconescu, T. Ohta, G. L. Kellogg, J. T. Robinson and I. V. Vlassiouk, Phys. Rev. B **87** (4), 041406 (2013).
37. P. V. C. Medeiros, S. Stafström and J. Björk, Phys. Rev. B **89** (4), 041407 (2014).
38. P. V. C. Medeiros, S. S. Tsirkin, S. Stafström, and J. Björk, Phys. Rev. B 91 041116 2015
39. T. B. Boykin and G. Klimeck, Phys. Rev. B **71** (11), 115215 (2005).
40. T. B. Boykin, N. Kharche, G. Klimeck and M. Korkusinski, J. Phys.: Cond. Matt. **19** (3), 036203 (2007).
41. M. Tomić, H. O. Jeschke and R. Valentí, Phys. Rev. B **90** (19), 195121 (2014).
42. E. P. O'Reilly, A. Lindsay and S. Fahy, J. Phys. Cond. Matt. **16** (31), S3257-S3276 (2004).
43. V. Lordi, H. B. Yuen, S. R. Bank and J. S. Harris, Appl. Phys. Lett. **85** (6), 902-904 (2004).
44. P. R. C. Kent and A. Zunger. Phys. Rev. B 64, 115208 (2001)